%% file: text.tex
\documentstyle[12pt,graphicx]{article}
\begin{document}
\baselineskip 0.6cm

\def\simgt{\mathrel{\lower2.5pt\vbox{\lineskip=0pt\baselineskip=0pt
           \hbox{$>$}\hbox{$\sim$}}}}
\def\simlt{\mathrel{\lower2.5pt\vbox{\lineskip=0pt\baselineskip=0pt
           \hbox{$<$}\hbox{$\sim$}}}}

\def\drawbox#1#2{\hrule height#2pt
    \hbox{\vrule width#2pt height#1pt \kern#1pt \vrule width#2pt}
  \hrule height#2pt}
\def\Fund#1#2{\vcenter{\vbox{\drawbox{#1}{#2}}}}
\def\Asym#1#2{\vcenter{\vbox{\drawbox{#1}{#2} \kern-#2pt \drawbox{#1}{#2}}}}
\def\fund{\Fund{6.5}{0.4}}
\def\asym{\Asym{6.5}{0.4}}

\begin{titlepage}

\begin{flushright}
UCB-PTH-07/18 \\
\end{flushright}

\vskip 2.4cm

\begin{center}

{\Large \bf 
A Simple and Realistic Model of Supersymmetry Breaking
}

\vskip 1.0cm

{\large Yasunori Nomura$^{a,b}$ and Michele Papucci$^{c}$}

\vskip 0.4cm

$^a$ {\it Department of Physics, University of California,
          Berkeley, CA 94720} \\
$^b$ {\it Theoretical Physics Group, Lawrence Berkeley National Laboratory,
          Berkeley, CA 94720} \\
$^c$ {\it School of Natural Sciences, Institute for Advanced Study,
          Princeton, NJ 08540}

\vskip 1.2cm

\abstract{We present a simple and realistic model of supersymmetry 
 breaking.  In addition to the minimal supersymmetric standard model, 
 we only introduce a hidden sector gauge group $SU(5)$ and three 
 fields $X$, $F$ and $\bar{F}$.  Supersymmetry is broken at a local 
 minimum of the potential, and its effects are transmitted to the 
 supersymmetric standard model sector through both standard model 
 gauge loops and local operators suppressed by the cutoff scale, which 
 is taken to be the unification scale.  The form of the local operators 
 is controlled by a $U(1)$ symmetry.  The generated supersymmetry 
 breaking and $\mu$ parameters are comparable in size, and no flavor 
 or $CP$ violating terms arise.  The spectrum of the first two generation 
 superparticles is that of minimal gauge mediation with the number of 
 messengers $N_{\rm mess} = 5$ and the messenger scale $10^{11}~{\rm GeV} 
 \simlt M_{\rm mess} \simlt 10^{13}~{\rm GeV}$.  The spectrum of the 
 Higgs bosons and third generation superparticles, however, can deviate 
 from it.  The lightest supersymmetric particle is the gravitino with 
 a mass of order $(1~{\rm -}~10)~{\rm GeV}$.}

\end{center}
\end{titlepage}

\section{Introduction}
\label{sec:intro}

Weak scale supersymmetry has long been the leading candidate for physics 
beyond the standard model.  It not only stabilizes the Higgs potential 
against potentially large radiative corrections, but also provides 
a successful prediction for the weak mixing angle through gauge coupling 
unification~\cite{Dimopoulos:1981zb}.  This framework, however, also 
introduces several new mysteries.  Chief amongst these are:
\begin{itemize}
\item
What is the origin of supersymmetry breaking, whose scale is 
hierarchically smaller than the Planck scale?
\item
Why is the supersymmetric mass for the Higgs doublets ($\mu$ parameter) 
the same order of magnitude as the supersymmetry breaking masses?
\item
Why have we not already observed flavor changing or $CP$ violating 
processes that are expected to occur in generic weak scale supersymmetric 
theories?
\item
Why does the proton not decay very rapidly through the processes 
that are allowed in general supersymmetric theories?
\end{itemize}
In this paper we present a simple and realistic model of weak scale 
supersymmetry which addresses these questions.

The model we consider is very simple.  In addition to the minimal 
supersymmetric standard model (MSSM), we only introduce a hidden 
sector gauge group $G_{\rm hid} = SU(5)_{\rm hid}$ and three fields 
$X$, $F$ and $\bar{F}$.  The quantum numbers of these fields under 
$SU(5)_{\rm hid} \times G_{\rm SM}$, where $G_{\rm SM}$ is the standard 
model gauge group, are $X({\bf 1},{\bf 1})$, $F({\bf 5},{\bf 5}^*)$ 
and $\bar{F}({\bf 5}^*,{\bf 5})$.  Here, we have used the language 
of $SU(5)_{\rm SM} \supset G_{\rm SM}$ for simplicity, although 
$G_{\rm SM}$ does not have to be unified into a single gauge group. 
The model has the superpotential interaction
\begin{equation}
  W = \lambda X F \bar{F},
\label{eq:W-anom}
\end{equation}
as well as the K\"ahler potential interactions $K = -|X|^4/4M_*^2 
+ (X^\dagger H_u H_d/M_* + {\rm h.c.})$, where $H_u$ and $H_d$ 
are the two Higgs doublets of the MSSM, and $O(1)$ coefficients 
are omitted.  The scale $M_*$ suppressing the K\"ahler potential 
interactions (the effective cutoff scale) is taken to be the 
unification scale, $M_* \simeq 10^{16}~{\rm GeV}$.  We find that 
this simple structure, together with a $U(1)$ symmetry controlling 
the form of the interactions, is essentially all we need to 
address the questions listed above.

The scale of supersymmetry breaking in our model is generated 
dynamically~\cite{Witten:1981nf}.  With the interaction of 
Eq.~(\ref{eq:W-anom}), the dynamics of $G_{\rm hid}$ generates 
the effective superpotential $W_{\rm eff} = \lambda X \Lambda^2$ 
for $\lambda X \simgt \Lambda$, where $\Lambda$ is the dynamical 
scale of $G_{\rm hid}$~\cite{Dimopoulos:1997ww,Murayama:1997pb}. 
This breaks supersymmetry on a plateau of the potential at 
$X \simgt \Lambda/\lambda$, with the scale of supersymmetry 
breaking given by $|F_X|^{1/2} = \lambda^{1/2} \Lambda$.  The 
field $X$ is a pseudo-flat direction, and we consider stabilizing 
it using supergravity effects.  In particular, we consider the 
$X$ mass term arising from the higher order K\"ahler potential 
term, $-|X|^4/4M_*^2$, and the $X$ linear term appearing in 
supergravity~\cite{Hall:1983iz,Kitano:2006wz}.  This gives 
$\langle X \rangle \approx M_*^2/M_{\rm Pl}$, where $M_{\rm Pl} 
\simeq 10^{18}~{\rm GeV}$ is the reduced Planck scale.  With this 
value of $\langle X \rangle$, the $\mu$ parameter arising from 
the K\"ahler potential~\cite{Giudice:1988yz} and the supersymmetry 
breaking masses arising from integrating out the $F,\bar{F}$ 
fields~\cite{Dine:1981gu,Dine:1994vc} are comparable~\cite{Ibe:2007km}. 
No flavor violating or $CP$ violating effects arise.  The particular 
form of the superpotential and the K\"ahler potential of the model 
is enforced by a global $U(1)$ symmetry under which $X$, $F$, 
$\bar{F}$, $H_u$ and $H_d$ carry the charges of $2$, $-1$, $-1$, 
$1$ and $1$, respectively.  (The charges of the matter fields are 
chosen accordingly.)  This symmetry is crucial to control the size 
of the $\mu$ parameter.  The same $U(1)$ symmetry can also be used 
to forbid dangerous operators leading to rapid proton decay.

The model provides a complete description of the dynamics, including 
supersymmetry breaking and its mediation, below the effective cutoff 
scale of $M_* \simeq 10^{16}~{\rm GeV}$.  We find that the coupling 
$\lambda$ in Eq.~(\ref{eq:W-anom}) must be in the range $10^{-6} 
\simlt \lambda^2 \simlt 10^{-3}$.  This and other properties of the 
model lead to several testable predictions.  In particular,
\begin{itemize}
\item
The spectrum of the first two generation squarks and sleptons 
is that of minimal gauge mediation with the number of messengers 
$N_{\rm mess} = 5$ and the messenger scale $10^{11}~{\rm GeV} 
\simlt M_{\rm mess} \simlt 10^{13}~{\rm GeV}$.%
\footnote{After submitting this paper, Ref.~\cite{Ibe:2007gf} appeared 
 which claims that the messenger scale must be smaller than about 
 $10^{10}~{\rm GeV}$.  We disagree with this.  By choosing $M_* \simeq 
 2 \times 10^{16}~{\rm GeV}$ and $\lambda \simeq 5 \times 10^{-3}$, 
 for example, we can obtain a realistic phenomenology with $M_{\rm mess} 
 \simeq 10^{12}~{\rm GeV}$, as can be seen from the analysis in 
 sections~\ref{sec:model} and \ref{sec:masses}.  The paper~\cite{Ibe:2007gf} 
 also studied the range $M_{\rm mess} \simlt 10^{10}~{\rm GeV}$, which 
 is outside the regime of validity of our present analysis, and found 
 a consistent supersymmetry breaking minimum in a branch of the moduli 
 space.  Overall, the range of the messenger scale in the present model 
 is, then, $10^{5}~{\rm GeV} \simlt M_{\rm mess} \simlt 10^{13}~{\rm 
 GeV}$ (except possibly for values around $M_{\rm mess} \approx 
 (10^{10}$~--~$10^{11})~{\rm GeV}$, where there is no theoretical 
 control over the dynamics).}
\item
The Higgs soft masses, $m_{H_u}^2$ and $m_{H_d}^2$, and the third 
generation squark and slepton masses deviate from those of minimal 
gauge mediation, but the deviations are parameterized by two free 
parameters.
\item
The lightest supersymmetric particle is the gravitino with a mass 
of order $(1~{\rm -}~10)~{\rm GeV}$.
\end{itemize}
We also find that the model presented here can be naturally ``derived'' 
by making a series of simple hypotheses for solutions to the issues 
listed at the beginning.  We consider that the existence of such 
an argument, together with the simplicity of its structure, makes 
the model very attractive.

In the next section, we provide this argument.  The actual 
model is presented and analyzed in the following three sections: 
sections~\ref{sec:model}, \ref{sec:masses} and \ref{sec:eft}. 
The $U(1)$ symmetry of the model is elaborated further in 
section~\ref{sec:pheno}.  Finally, a summary and discussions 
are given in section~\ref{sec:summary}, where the robustness 
of the predictions under possible modifications of the model 
is discussed.

\section{``Derivation''}
\label{sec:deriv}

One of the most promising ways to address why the $\mu$ parameter is 
the same order as the supersymmetry breaking masses is to forbid $\mu$ 
in the supersymmetric limit and generate it through supersymmetry 
breaking.  There are two classes of symmetries that can achieve this. 
One is an $R$ symmetry under which the Higgs bilinear $H_u H_d$ is 
neutral.  The other is a non-$R$ symmetry under which $H_u H_d$ is 
charged.  (An $R$ symmetry under which $H_u H_d$ carries a nonzero 
charge other than $2$ also falls in this latter class.)  In these 
cases, $\mu$ can be generated by coupling $H_u H_d$ to the supersymmetry 
breaking superfield $X = \theta^2 F_X$ in the K\"ahler potential
\begin{equation}
  K = \frac{1}{M_*} X^\dagger H_u H_d + {\rm h.c.},
\label{eq:K-mu}
\end{equation}
where $M_*$ is some mass scale~\cite{Giudice:1988yz}.  The $X$ field 
is neutral in the former case, while it has the same nonzero charge 
as $H_u H_d$ in the latter case.  An interesting point is that in 
either case the symmetry that forbids a (potentially large) Higgs 
mass in the supersymmetric limit also forbids a linear $X$ term in 
the superpotential, $W = M^2 X$, which contributes to supersymmetry 
breaking with the (potentially large) breaking scale, $M$.  The 
reason why the Higgs doublet mass is not the Planck scale is related 
to the reason why supersymmetry is not broken at the Planck scale.

Let us now adopt the latter case: a non-$R$ $U(1)$ global symmetry, 
$U(1)_H$, forbidding the $H_u H_d$ and $X$ terms in the superpotential. 
This has an advantage that the K\"ahler potential term $K = X^\dagger 
X H_u H_d/M_*^2 + {\rm h.c.}$ is not allowed, so that the holomorphic 
supersymmetry breaking Higgs mass-squared ($B\mu$ parameter) is 
not generated at the same order as $\mu^2$.  This avoids generating 
problematic large $CP$ violation at low energies, such as an electron 
electric dipole moment beyond the current experimental limit, which 
would arise if both $\mu$ and $B\mu$ were generated with a comparable 
order and arbitrary complex phases.%
\footnote{For $M_* \approx M_{\rm Pl}$, $B\mu \approx F_X^2/M_* 
 M_{\rm Pl}$ generated by gravity mediation can be comparable to 
 $\mu^2 \approx F_X^2/M_*^2$.  This contribution, however, does 
 not necessarily introduce a new $CP$ violating phase.  Moreover, 
 the generated $B\mu$ is much smaller than $\mu^2$ for $M_* \ll 
 M_{\rm Pl}$, which is the region we are interested in; see below.}
The question then is how to generate a linear $X$ term in the 
superpotential, needed to break supersymmetry, and a coupling of 
$X$ to the standard model gauge supermultiplets, needed to generate 
the gaugino masses in the MSSM.  These operators are both forbidden 
by the $U(1)_H$ symmetry.

It is immediately clear that if $U(1)_H$ has anomalies with respect to 
the hidden sector gauge group $G_{\rm hid}$ and the standard model gauge 
group $G_{\rm SM}$, the low energy effective theory has the operators
\begin{equation}
  {\cal L} = \int\!d^2\theta\, \frac{A_{\rm hid}}{32\pi^2}\, 
    (\ln X)\, {\cal W}^\alpha {\cal W}_\alpha + {\rm h.c.},
\label{eq:anom-hid}
\end{equation}
and
\begin{equation}
  {\cal L} = \sum_a \int\!d^2\theta\, \frac{A_{\rm SM}}{32\pi^2}\, 
    (\ln X)\, {\cal W}^{a\alpha} {\cal W}^a_\alpha + {\rm h.c.},
\label{eq:anom-SM}
\end{equation}
respectively, where ${\cal W}_\alpha$ and ${\cal W}^a_\alpha$ $(a=1,2,3)$ 
are the field-strength superfields for $G_{\rm hid}$ and $G_{\rm SM}$,%
\footnote{The field-strength superfields are normalized such that the 
 gauge kinetic terms are given by ${\cal L}_{\rm kin} = \int\!d^2\theta\, 
 \{ (1/4g^2) {\cal W}^\alpha {\cal W}_\alpha + \sum_a (1/4g_a^2) 
 {\cal W}^{a\alpha} {\cal W}^a_\alpha \} + {\rm h.c.}$, where $g$ 
 and $g_a$ are the gauge couplings for $G_{\rm hid}$ and $G_{\rm SM}$.}
and $A_{\rm hid}$ and $A_{\rm SM}$ the $U(1)_H$-$G_{\rm hid}^2$ and 
$U(1)_H$-$G_{\rm SM}^2$ anomalies.  Here, we have normalized the 
$U(1)_H$ charge of $X$ to be $2$, and assumed a nonvanishing vacuum 
expectation value (VEV) of $X$, as well as the universality of $A_{\rm SM}$ 
with respect to the three gauge group factors of $G_{\rm SM}$.  Note 
that the form of Eqs.~(\ref{eq:anom-hid},~\ref{eq:anom-SM}) is completely 
dictated by the anomalous $U(1)_H$ symmetry: under $X \rightarrow 
e^{2i\alpha} X$, the Lagrangian must transform as ${\cal L} \rightarrow 
{\cal L} - (\alpha/32\pi^2)(A_{\rm hid} F_{\mu\nu} \tilde{F}^{\mu\nu} 
+ A_{\rm SM} F^a_{\mu\nu} \tilde{F}^{a\mu\nu})$, where $F_{\mu\nu}$ and 
$F^a_{\mu\nu}$ are the field strengths for $G_{\rm hid}$ and $G_{\rm SM}$. 
We find that the operator of Eq.~(\ref{eq:anom-hid}) is the one needed 
to generate a superpotential for $X$ through the hidden sector dynamics, 
and the operators of Eq.~(\ref{eq:anom-SM}) are responsible for the 
gaugino masses.

The $U(1)_H$ anomalies with $G_{\rm hid}$ and $G_{\rm SM}$ can be enforced 
by coupling $X$ to a vector-like field(s) $F, \bar{F}$: $W = \lambda 
X F \bar{F}$ in Eq.~(\ref{eq:W-anom}).  If $F$ and $\bar{F}$ are charged 
under $G_{\rm hid}$ ($G_{\rm SM}$), a nonvanishing $U(1)_H$-$G_{\rm hid}^2$ 
($U(1)_H$-$G_{\rm SM}^2$) anomaly results: $A_{\rm hid} = -2 T_F^{\rm hid}$ 
($A_{\rm SM} = -2 T_F^{\rm SM}$), where $T_F^{\rm hid}$ ($T_F^{\rm SM}$) 
is the Dynkin index of $F$ under $G_{\rm hid}$ ($G_{\rm SM}$), normalized 
to be $1/2$ for the fundamental representation of $SU(N)$.  The operators 
of Eqs.~(\ref{eq:anom-hid},~\ref{eq:anom-SM}) then appear after integrating 
out the $F,\bar{F}$ fields.  With the interaction Eq.~(\ref{eq:anom-hid}), 
the dynamics of $G_{\rm hid}$ generates the effective superpotential
\begin{equation}
  W_{\rm eff} = \Lambda_{\rm eff}^3 
    = (\lambda X)^{T_F^{\rm hid}/T_G^{\rm hid}} 
      \Lambda^{(3T_G^{\rm hid}-T_F^{\rm hid})/T_G^{\rm hid}},
\label{eq:W-eff}
\end{equation}
through gaugino condensation~\cite{Veneziano:1982ah}.  Here, 
$\Lambda_{\rm eff}$ and $\Lambda$ are the dynamical scales of the 
low-energy effective pure $G_{\rm hid}$ theory and the original 
$G_{\rm hid}$ theory, respectively, and $T_G^{\rm hid}$ is the 
Dynkin index for the adjoint representation of $G_{\rm hid}$ 
($N$ for $G_{\rm hid} = SU(N)$).  We find that for $T_F^{\rm hid} 
= T_G^{\rm hid}$ the generated superpotential is linear in $X$, 
so that supersymmetry is broken by $\partial W_{\rm eff}/\partial 
X \neq 0$.

Supersymmetry breaking by the operator of Eq.~(\ref{eq:W-eff}) 
was considered in Refs.~\cite{Dimopoulos:1997ww,Murayama:1997pb} 
to build models of ``direct'' gauge mediation.  With $T_F^{\rm hid} 
= T_G^{\rm hid}$, $X$ is a pseudo-flat direction and has a supersymmetry 
breaking plateau at $X \simgt \Lambda/\lambda$.  (For $X \simlt 
\Lambda/\lambda$ the mass of $F, \bar{F}$ becomes smaller than the 
dynamical scale $\Lambda$, and our present analysis breaks down.) 
Now, suppose that $X$ is stabilized at some value $\langle X \rangle$. 
Integrating out the $F,\bar{F}$ fields charged under $G_{\rm SM}$, 
then, generates gauge mediated contributions to the MSSM gaugino 
and scalar masses of order
\begin{equation}
  m_{\rm GMSB} \approx 
    \frac{g^2}{16\pi^2} \frac{\lambda \Lambda^2}{\langle X \rangle},
\label{eq:gmsb}
\end{equation}
where $g$ represents the standard model gauge couplings~\cite{Dine:1981gu,%
Dine:1994vc}.  On the other hand, the $\mu$ parameter generated by 
Eq.~(\ref{eq:K-mu}) is of order
\begin{equation}
  \mu \approx \frac{\lambda \Lambda^2}{M_*}.
\label{eq:mu}
\end{equation}
How can these two contributions be the same order?  In the absence of 
other sources for the $\mu$ or supersymmetry breaking parameters, these 
two contributions must be comparable, which requires $\langle X \rangle 
\approx (g^2/16\pi^2)M_* \approx 10^{-2} M_*$.  Is there any reason 
to expect $\langle X \rangle$ to be in this particular range?

In fact, one of the simplest ways to stabilize $X$ gives us such 
a reason.  The existence of the operator Eq.~(\ref{eq:K-mu}) suggests 
that the $X$ field also has higher dimension operators in the K\"ahler 
potential suppressed by powers of $M_*$.  Now, suppose that the 
coefficient of the lowest such operator, $|X|^4$, is negative:
\begin{equation}
  K = - \frac{1}{4 M_*^2} |X|^4,
\label{eq:X4}
\end{equation}
where we have put the factor $1/4$ to take into account the 
symmetry factor.  This gives a positive mass squared to the pseudo-flat 
direction $X$, since the potential has the contribution $V \sim 
|\partial W/\partial X|^2 (\partial^2 K/\partial X^\dagger \partial 
X)^{-1} \supset (|\lambda \Lambda^2|^2/M_*^2) |X|^2$.  The resulting 
minimum, however, is not at the origin (which would push $\langle 
X \rangle$ away from the plateau).  This is because in supergravity 
the potential receives the contribution $-3|W|^2/M_{\rm Pl}^2 
- \{ F_X (\partial K/\partial X) W + {\rm h.c.} \}/M_{\rm Pl}^2$, 
leading to a linear term in $X$~\cite{Kitano:2006wz}: $V \sim 
\lambda \Lambda^2 c (X + X^\dagger)/M_{\rm Pl}^2$, where $c$ 
is the constant term in the superpotential needed to cancel the 
cosmological constant.  Balancing these two effects and setting 
$c \sim \lambda \Lambda^2 M_{\rm Pl}$ to cancel the positive 
vacuum energy of the plateau leads to
\begin{equation}
  \langle X \rangle \approx \frac{M_*^2}{M_{\rm Pl}}.
\label{eq:X-stab}
\end{equation}
This gives the desired relation $\langle X \rangle \approx 10^{-2} 
M_*$ if we choose $M_*$ to be the unification scale, $M_* \simeq 
10^{16}~{\rm GeV}$, one of the natural choices for the cutoff 
of the MSSM.  This coincidence of scales was noticed in 
Ref.~\cite{Ibe:2007km}, where the effective field theory 
description/parameterization of the class of dynamics under 
consideration was also discussed in detail.

To stabilize $X$ at the value of Eq.~(\ref{eq:X-stab}) and make the 
supersymmetry breaking masses and $\mu$ comparable, the dominant 
deformation of the plateau must come from the two effects described 
above.  This gives a restriction on the range of the parameter 
$\lambda$, appearing in Eq.~(\ref{eq:W-anom}).  First, the effect 
from $F,\bar{F}$ loops on the $X$ potential must be subdominant. 
This gives an upper bound on $\lambda$.  The stabilized value of 
$\langle X \rangle$ must also satisfy $M_F \equiv \lambda \langle 
X \rangle > \Lambda$, giving a lower bound on $\lambda$.  We study 
these and other bounds on $\lambda$, and find that there is a region 
where all the requirements are satisfied.  This, together with 
Eq.~(\ref{eq:X-stab}), is then translated into the allowed range 
for the mass of the $F,\bar{F}$ fields, $M_F$.

The arguments described above lead to the following picture 
for supersymmetric theories in which the questions listed in 
section~\ref{sec:intro} are addressed:
\begin{itemize}
\item
The effective cutoff scale of the MSSM is the unification scale, 
$M_* \approx 10^{16}~{\rm GeV}$.  The Higgs and supersymmetry breaking 
fields have higher dimension operators suppressed by powers of $M_*$. 
This is consistent with successful gauge coupling unification in 
supersymmetric theories.
\item
There is no tree-level operator connecting matter and supersymmetry 
breaking fields in the K\"ahler potential $K \approx M^\dagger M X^\dagger 
X/M_*^2$, where $M$ represents the MSSM matter fields.  Such operators 
would generically lead to large flavor changing neutral currents at 
low energies, and so should be suppressed.  This property must arise 
from the theory at or above the effective cutoff scale $M_*$.
\item
The supersymmetry breaking field $X$ is charged under $U(1)_H$ which 
has anomalies with respect to both $G_{\rm hid}$ and $G_{\rm SM}$. 
The anomalies are enforced by coupling $X$ to a vector-like pair(s) of 
fields $F$ and $\bar{F}$ in the superpotential, $W = \lambda X F \bar{F}$. 
The simplest possibility to make both the $U(1)_H$-$G_{\rm hid}^2$ and 
$U(1)_H$-$G_{\rm SM}^2$ anomalies nonvanishing is to consider $F,\bar{F}$ 
to be charged under both $G_{\rm hid}$ and $G_{\rm SM}$.
\item
To generate the effective superpotential linear in $X$ by the hidden 
sector dynamics, the Dynkin index of $F,\bar{F}$ under $G_{\rm hid}$ 
must be the same as that of the adjoint representation of $G_{\rm hid}$: 
$T_F^{\rm hid} = T_G^{\rm hid}$.  The simplest possibility to realize 
this is to consider that $G_{\rm hid} = SU(5)_{\rm hid}$, and that 
$F,\bar{F}$ are ``bi-fundamental'' under $SU(5)_{\rm hid} \times 
SU(5)_{\rm SM}$: $F({\bf 5},{\bf 5}^*)$ and $\bar{F}({\bf 5}^*,{\bf 5})$, 
where $SU(5)_{\rm SM} \supset G_{\rm SM}$.
\end{itemize}
In the next two sections, we present an explicit model based on these 
observations.  The analysis of the dynamics of the model using effective 
field theory will be given in section~\ref{sec:eft}.

\section{Model}
\label{sec:model}

The gauge group of the model is $SU(5)_{\rm hid} \times G_{\rm SM}$. 
In addition to the MSSM fields, $Q$, $U$, $D$, $L$, $E$, $H_u$ and 
$H_d$, which are all singlet under $SU(5)_{\rm hid}$, we introduce 
a singlet field and a pair of ``bi-fundamental'' fields:
\begin{equation}
  X({\bf 1},{\bf 1}),
\qquad
  F({\bf 5},{\bf 5}^*),
\qquad
  \bar{F}({\bf 5}^*,{\bf 5}),
\label{eq:matter-content}
\end{equation}
where the numbers in parentheses represent quantum numbers under 
$SU(5)_{\rm hid} \times SU(5)_{\rm SM}$.  Here, we have used the language 
of $SU(5)_{\rm SM} \supset G_{\rm SM}$ for simplicity of notation, but 
$G_{\rm SM}$ does not have to be unified into a single gauge group.

We now introduce a (anomalous) $U(1)$ global symmetry, $U(1)_H$, under 
which the fields transform as
\begin{equation}
  X(2), \quad F(-1), \quad \bar{F}(-1),
\label{eq:H-XFF}
\end{equation}
\begin{equation}
  Q(x), \quad U(-1-x), \quad D(-1-x), \quad L(y), \quad E(-1-y),
\label{eq:H-matter}
\end{equation}
\begin{equation}
  H_u(1), \quad H_d(1),
\label{eq:H-Higgs}
\end{equation}
where $x$ and $y$ are some numbers.  The charges of $H_u$ and $H_d$ are 
determined so that the term $X^\dagger H_u H_d$ is allowed in the K\"ahler 
potential (we can use a hypercharge rotation to make the $H_u$ and $H_d$ 
charges equal without loss of generality), and those of matter are 
determined so that the Yukawa couplings are invariant under $U(1)_H$.

We take the cutoff scale of our theory to be around the unification 
scale, $M_* \simeq 10^{16}~{\rm GeV}$.  This preserves successful 
gauge coupling unification.  The most general K\"ahler potential and 
superpotential among the $X$, $F$ and $\bar{F}$ fields consistent 
with $SU(5)_{\rm hid} \times G_{\rm SM} \times U(1)_H$ are then
\begin{equation}
  K = K_{\rm kin} - \frac{1}{4 M_*^2} (X^\dagger X)^2 + \cdots,
\label{eq:K_X}
\end{equation}
\begin{equation}
  W = c + \lambda X F \bar{F} + \cdots,
\label{eq:W_X}
\end{equation}
where $K_{\rm kin}$ represents the canonically normalized kinetic terms, 
$c$ is a constant term in the superpotential, and $\lambda$ is a coupling 
constant.  Here, we have assumed that the coefficient of the second term 
in Eq.~(\ref{eq:K_X}) is negative, and absorbed its magnitude into the 
definition of $M_*$.  The parameters $c$ and $\lambda$ are taken to be 
real and positive without loss of generality by using $U(1)_R$ and 
$F\bar{F}$ rotations.  The most general interactions between $X$, $F$, 
$\bar{F}$ and the Higgs fields are
\begin{equation}
  K \approx \left( \frac{1}{M_*} X^\dagger H_u H_d + {\rm h.c.} \right) 
    + \frac{1}{M_*^2} X^\dagger X H_u^\dagger H_u 
    + \frac{1}{M_*^2} X^\dagger X H_d^\dagger H_d + \cdots,
\label{eq:K_H-X}
\end{equation}
\begin{equation}
  W = \frac{\eta}{M_*} F \bar{F} H_u H_d + \cdots,
\label{eq:W_H-X}
\end{equation}
where $\eta$ is a dimensionless coupling.  Note that we have taken the 
theory to be weakly coupled at $M_*$ (or strongly coupled at $\approx 
4\pi M_*$), so that the dimensionless coefficients in the K\"ahler 
potential, omitted in Eq.~(\ref{eq:K_H-X}), are naturally of order 
unity.  On the other hand, the superpotential couplings can be naturally 
smaller because they are radiatively stable.  We assume the absence of 
interactions between the $X$ and matter fields suppressed by powers of 
$M_*$, as stated in the previous section.  This can be achieved, for 
example, if the $X$ and matter fields are localized to distant points 
in (small) extra dimensions, with the Higgs fields propagating in 
the bulk.

We now demonstrate that the simple model described above gives 
successful supersymmetry breaking and its mediation.  We first 
consider the minimization of the potential for $X$.  At low energies, 
the fields $F$ and $\bar{F}$ decouple at the mass $M_F = \lambda X$, 
and the gaugino condensation of $SU(5)_{\rm hid}$ generates the 
superpotential $W = \Lambda_{\rm eff}^3$.  Here, $\Lambda_{\rm eff}$ 
is the effective dynamical scale for the low-energy pure $SU(5)$ 
gauge theory, which is related to the dynamical scale $\Lambda$ 
of the original $SU(5)_{\rm hid}$ by the matching condition 
$\Lambda_{\rm eff}^3 = M_F \Lambda^2 = \lambda X \Lambda^2$.%
\footnote{The precise definition of $\Lambda$ here is given such 
 that the generated superpotential is $W = \Lambda_{\rm eff}^3$. 
 \label{ft:Lambda}}
This implies that the dynamics of $SU(5)_{\rm hid}$ generates 
the superpotential
\begin{equation}
  W_{\rm eff} = \lambda \Lambda^2 X,
\label{eq:W_eff}
\end{equation}
which leads to a supersymmetry breaking plateau for the $X$ potential, 
taking the form $V \sim |\lambda \Lambda^2|^2$ in the limit that 
we neglect supergravity and higher order corrections.  Note that 
this analysis is valid only for $|M_F| = |\lambda X| \simgt \Lambda$. 
For $|\lambda X| \simlt \Lambda$, $SU(5)_{\rm hid}$ has 5~flavors of 
light ``quarks,'' and there are supersymmetric minima at $X = 0$ even 
at the quantum level.%
\footnote{The minima are at $X = 0$, ${\rm tr}(M^i_j) = 0$ and 
 ${\rm det}(M^i_j)  - B\bar{B} = (\Lambda^2/5)^5$, where $M^i_j 
 \equiv F^i_\alpha \bar{F}^\alpha_j$, $B \equiv \epsilon^{\alpha 
 \beta \gamma \delta \eta} F^i_\alpha F^j_\beta F^k_\gamma F^l_\delta 
 F^m_\eta \epsilon_{ijklm}$ and $\bar{B} \equiv \epsilon_{\alpha \beta 
 \gamma \delta \eta} \bar{F}^\alpha_i \bar{F}^\beta_j \bar{F}^\gamma_k 
 \bar{F}^\delta_l \bar{F}^\eta_m \epsilon^{ijklm}$ are the ``meson,'' 
 ``baryon'' and ``antibaryon'' superfields of $SU(5)_{\rm hid}$, with 
 $\alpha,\beta,\cdots$ and $i,j,\cdots$ representing the indices 
 of $SU(5)_{\rm hid}$ and $SU(5)_{\rm SM}$, respectively.}

The supersymmetry breaking plateau at $|X| \simgt \Lambda/\lambda$ is 
distorted by a number of corrections, including effects from higher 
dimension operators in Eq.~(\ref{eq:K_X}), supergravity terms, and 
loops of the $F$ and $\bar{F}$ fields.  With the negative sign for its 
coefficient, the second term in Eq.~(\ref{eq:K_X}) induces a positive 
mass-squared term for $X$ in the potential, $\delta V = (\lambda^2 
\Lambda^4/M_*^2) |X|^2$.  On the other hand, supergravity corrections 
lead to a linear term in $X$, $\delta V = -2 \lambda \Lambda^2 c 
(X + X^\dagger)/M_{\rm Pl}^2$.%
\footnote{In the superconformal calculus formulation of 
 supergravity~\cite{Cremmer:1978hn}, the $X$ linear term in the scalar 
 potential arises from the $F$-term VEV, $F_\varphi = c/M_{\rm Pl}^2$, 
 of the chiral compensator field $\varphi$ through the superpotential 
 term $W = \varphi^2 \lambda X \Lambda^2$.  From the viewpoint 
 of the original theory Eqs.~(\ref{eq:K_X}~--~\ref{eq:W_H-X}), 
 this effect is understood to arise from the anomaly-mediated 
 contribution~\cite{Randall:1998uk} to the $SU(5)_{\rm hid}$ 
 gaugino mass.}
These two effects lead to a minimum at $\langle X \rangle = 2 c 
M_*^2/\lambda \Lambda^2 M_{\rm Pl}^2$.  The value of $c$ is determined 
by the condition of a vanishing cosmological constant at this minimum 
in the supergravity potential, $V \simeq \lambda^2 \Lambda^4 - 3 
c^2/M_{\rm Pl}^2 = 0$.  This leads to $c \simeq \lambda \Lambda^2 
M_{\rm Pl}/\sqrt{3}$, and hence
\begin{equation}
  \langle X \rangle \simeq \frac{2 M_*^2}{\sqrt{3} M_{\rm Pl}} 
    \approx 10^{14}~{\rm GeV},
\qquad
  F_X \simeq -\lambda \Lambda^2,
\label{eq:X-FX-VEV}
\end{equation}
where $F_X \equiv \langle -\partial W^\dagger/\partial X^\dagger 
- (\partial K/\partial X^\dagger) W^\dagger/M_{\rm Pl}^2 \rangle$ 
is the supersymmetry breaking VEV for the $X$ superfield.  Note that 
$\langle X \rangle \approx M_*(M_*/M_{\rm Pl}) < M_*$, so that it 
stays within the regime where the effective field theory below $M_*$ 
is applicable.

The loops of the $F$, $\bar{F}$ fields also affect the $X$ potential. 
One such effect arises from the 1-loop Coleman-Weinberg correction to 
the K\"ahler potential $\delta K_{\rm CW} \simeq -(\lambda^2 n_F/16\pi^2) 
|X|^2 \ln(|\lambda X|^2/\mu_R^2)$, where $n_F \equiv n_G^2 = 25$ is the 
number of degrees of freedom for $F$, and $\mu_R$ the renormalization 
scale.  This effect is small enough not to destabilize the minimum 
of Eq.~(\ref{eq:X-FX-VEV}), as long as
\begin{equation}
  \frac{\lambda^2 n_G^2}{16\pi^2} \simlt 
    \left(\frac{M_*}{M_{\rm Pl}}\right)^2 
    \approx 10^{-4},
\label{eq:stab-cond}
\end{equation}
where $\lambda$ is evaluated at the scale $\approx \lambda \langle X 
\rangle$.  Another effect arises from the generation of the K\"ahler 
potential operator $\delta K_{{\cal W}^4} \approx (1/64)(N_{\rm SM}/16\pi^2)
|{\cal W}^\alpha {\cal W}_\alpha|^2/|\lambda X|^4$ with ${\cal W}^\alpha 
{\cal W}_\alpha \rightarrow 32\pi^2 \Lambda_{\rm eff}^3/N_{\rm hid} 
= 32\pi^2 \lambda X \Lambda^2/N_{\rm hid}$, where ${\cal W}_\alpha$ 
is the field-strength superfield for $SU(5)_{\rm hid}$, and $N_{\rm hid} 
= N_{\rm SM} = 5$ are the number of ``colors'' for $SU(5)_{\rm hid}$ 
and $SU(5)_{\rm SM}$ ($1/64$ is the symmetry factor).  This effect 
is unimportant as long as
\begin{equation}
  \lambda^2 \simgt \frac{\pi M_{\rm Pl}^3 |F_X|}{\sqrt{n_G} M_*^5},
\label{eq:stab-cond-2}
\end{equation}
where $n_G = 5$.  There is also a correction arising from nonperturbative 
effects of $SU(5)_{\rm hid}$, given by $\delta K_{\rm np} \approx k 
|\Lambda_{\rm eff}|^2 \approx k |\lambda \Lambda^2 X|^{2/3}$, where $k 
\approx (N_{\rm hid}^2/16\pi^2)^{1/3} \simeq O(1)$~\cite{Randall:1998ra}. 
This correction is irrelevant if
\begin{equation}
  \frac{|F_X| M_{\rm Pl}^5}{M_*^7} \simlt 10^2,
\label{eq:stab-cond-3}
\end{equation}
where we have used Eq.~(\ref{eq:X-FX-VEV}).  We assume that these 
conditions are satisfied, so that the minimum of $X$ is given 
by Eq.~(\ref{eq:X-FX-VEV}).  There is one remaining condition on 
$\lambda$ which comes from the requirement that the minimum lies 
on the plateau, $|\langle X \rangle| \simgt (4\pi/\sqrt{N_{\rm SM} 
N_{\rm hid}}) \Lambda/\lambda$.  Here, we have included the 
factor of $4\pi/\sqrt{N_{\rm SM}}$ suggested by naive dimensional 
analysis~\cite{Luty:1997fk}, and the factor $1/\sqrt{N_{\rm hid}}$ 
arises from our definition of $\Lambda$ (see footnote~\ref{ft:Lambda}). 
This gives
\begin{equation}
  \lambda^3 \simgt \frac{16\pi^2 M_{\rm Pl}^2 |F_X|}{n_G^2 M_*^4}.
\label{eq:lambda-cond}
\end{equation}
One implication of these conditions, Eqs.~(\ref{eq:stab-cond}~--%
~\ref{eq:lambda-cond}), will be discussed later.

\begin{figure}
\begin{center}
  \input{figure.tex}
\caption{The schematic depiction of the $X$ potential.}
\label{fig:pot-X}
\end{center}
\end{figure}
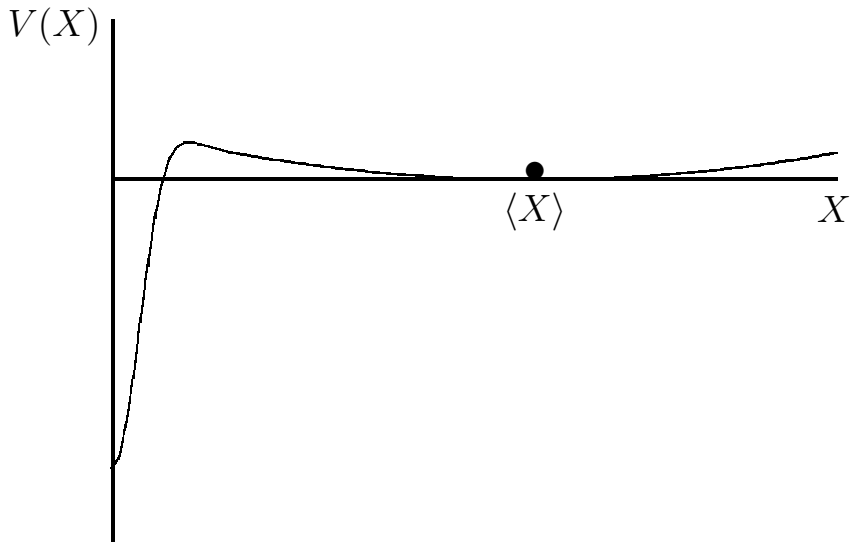

The potential for $X$ in our model is depicted schematically in 
Fig.~\ref{fig:pot-X}.
We are living in a very flat supersymmetry breaking plateau with 
$V \sim 0$.  The tunneling rate from our local minimum to the 
true (supersymmetric) minimum can be estimated using the technique 
of Ref.~\cite{Coleman:1977py}.  The decay rate per unit volume 
is given by $\Gamma/V \sim \langle X \rangle^4 e^{-B}$, where 
$B \sim 2\pi^2 \langle X \rangle^4/F_X^2 \sim 10^{20}$.  Here, 
we have used Eq.~(\ref{eq:FX-Ms}) below to obtain the numerical 
estimate for $B$.  We find that the lifetime of the local minimum 
is much larger than the age of the universe: $\Gamma/V \ll H_0^4$, 
where $H_0 \simeq 10^{-33}~{\rm eV}$ is the present Hubble constant. 
The mass squared for $X$ around the minimum is given by
\begin{equation}
  m_X^2 \simeq \left(\frac{F_X}{M_*}\right)^2.
\label{eq:mX2}
\end{equation}
As we will see, this is of the order of the weak scale squared.

\section{Superparticle Masses}
\label{sec:masses}

With the $X$ VEV and $F_X$ in Eq.~(\ref{eq:X-FX-VEV}), the supersymmetry 
breaking and $\mu$ parameters in the MSSM receive several contributions. 
First, the interactions in Eq.~(\ref{eq:K_H-X}) give
\begin{equation}
  \mu \approx \frac{F_X}{M_*},
\qquad
  m_{H_u}^2 \approx m_{H_d}^2 \approx \left(\frac{F_X}{M_*}\right)^2,
\label{eq:mu-mH2}
\end{equation}
at the scale $M_*$.  Here, $m_{H_u}^2$ and $m_{H_d}^2$ are the 
non-holomorphic supersymmetry breaking squared masses for $H_u$ 
and $H_d$.  On the other hand, the interaction of Eq.~(\ref{eq:W_H-X}) 
gives $\mu \approx \eta \Lambda^2/M_* \approx (\eta/\lambda) F_X/M_*$. 
(This is obtained by making the replacement $\lambda X \rightarrow 
\lambda X + \eta H_u H_d/M_*$ in Eq.~(\ref{eq:W_eff}).)  We assume 
$\eta \simlt \lambda$ so that this contribution is, at most, comparable 
to that in Eq.~(\ref{eq:mu-mH2}).%
\footnote{This is easily achieved, for example, if the suppressions 
 of $\lambda$ and $\eta$ have a common origin, such as the suppression 
 of the couplings between the $F$, $\bar{F}$ fields and the rest of 
 the fields (in which case we expect $\lambda \approx \eta$).}
The masses generated at the scale $M_*$ then take the form of 
Eq.~(\ref{eq:mu-mH2}), with the other supersymmetry breaking parameters  
essentially vanishing.  Note that here we have omitted $O(1)$ coefficients, 
so that the ratio of $m_{H_u}^2$ to $m_{H_d}^2$, for example, can be 
an arbitrary $O(1)$ number.

The supersymmetry breaking masses also receive contributions from loops 
of the $F$, $\bar{F}$ fields through gauge mediation, generated at the 
mass scale of these fields
\begin{equation}
  M_{\rm mess} \approx \lambda \langle X \rangle 
    \approx \frac{\lambda M_*^2}{M_{\rm Pl}}.
\label{eq:M_mess}
\end{equation}
The gaugino masses $M_a$ ($a=1,2,3$) and the scalar squared masses 
$m_{\tilde{f}}^2$ ($\tilde{f} = \tilde{q}, \tilde{u}, \tilde{d}, 
\tilde{l}, \tilde{e}$) receive contributions
\begin{equation}
  M_a = N_{\rm mess} \frac{g_a^2}{16\pi^2} \frac{F_X}{\langle X \rangle},
\label{eq:gauginos}
\end{equation}
\begin{equation}
  m_{\tilde{f}}^2 = 2 N_{\rm mess} \sum_a C_a^{\tilde{f}} 
      \left( \frac{g_a^2}{16\pi^2} \right)^2 
      \left| \frac{F_X}{\langle X \rangle} \right|^2,
\label{eq:scalars}
\end{equation}
where $a = 1,2,3$ represents the standard model gauge group factors, 
$g_a$ the standard model gauge couplings at $M_{\rm mess}$, and 
$C_a^{\tilde{f}}$ the group theory factors given by $(C_1^{\tilde{f}}, 
C_2^{\tilde{f}}, C_3^{\tilde{f}}) = (1/60,3/4,4/3)$, $(4/15,0,4/3)$, 
$(1/15,0,4/3)$, $(3/20,3/4,0)$ and $(3/5,0,0)$ for $\tilde{f} = \tilde{q}, 
\tilde{u}, \tilde{d}, \tilde{l}$ and $\tilde{e}$, respectively.  The 
contributions to $m_{H_u}^2$ and $m_{H_d}^2$ are the same as that 
to $m_{\tilde{l}}^2$.  The quantity $N_{\rm mess}$ is the number of 
messenger pairs, which is predicted as
\begin{equation}
  N_{\rm mess} = 5,
\label{eq:N_mess}
\end{equation}
in the present model.

The low-energy superparticle masses are obtained by evolving the 
parameters of Eq.~(\ref{eq:mu-mH2}) from $M_*$ to $M_{\rm mess}$, 
adding the contributions of Eqs.~(\ref{eq:gauginos},~\ref{eq:scalars}) 
at $M_{\rm mess}$, and then evolving the resulting parameters from 
$M_{\rm mess}$ down to the weak scale.  Note that since the gauge-mediated 
contributions of Eqs.~(\ref{eq:gauginos},~\ref{eq:scalars}) have the size
\begin{equation}
  M_a \approx (m_{\tilde{f}}^2)^{1/2} 
  \approx \frac{F_X}{M_*} 
    \left( \frac{g^2}{16\pi^2} \frac{M_{\rm Pl}}{M_*} \right) 
  \approx \frac{F_X}{M_*},
\label{eq:GMSB-size}
\end{equation}
where $g$ represents the standard model gauge couplings, they are 
comparable to the tree-level contributions to the Higgs-sector 
parameters of Eq.~(\ref{eq:mu-mH2}).%
\footnote{In contrast with the situation discussed in 
 Ref.~\cite{Ibe:2007km}, there is no reason in our theory why 
 the $\mu$ term must be suppressed compared with the gauge-mediated 
 contributions.  In fact, they are naturally expected to be 
 comparable.}
Setting the size of these contributions to be the weak scale
\begin{equation}
  \left|\frac{F_X}{M_*}\right| 
    \approx (100~{\rm GeV}~{\rm -}~1~{\rm TeV}),
\label{eq:FX-Ms}
\end{equation}
the value of $F_X$ is determined as $\sqrt{F_X} \approx 
(10^{9}$~--~$10^{9.5})~{\rm GeV}$.  With these values of $F_X$, the 
condition of Eq.~(\ref{eq:stab-cond-3}) is satisfied.  The messenger 
scale of gauge mediation, $M_{\rm mess}$, is given by Eq.~(\ref{eq:M_mess}), 
and is subject to the bounds on $\lambda$ in Eqs.~(\ref{eq:stab-cond}), 
(\ref{eq:stab-cond-2}) and (\ref{eq:lambda-cond}), which can be 
written as $\lambda^2 \simlt 10^{-3}$, $\lambda^2 \simgt 10^{-8}$ 
and $\lambda^3 \simgt 10^{-9}$ using Eq.~(\ref{eq:FX-Ms}).  These 
lead to the following range on the messenger scale of gauge mediation:
\begin{equation}
  10^{11}~{\rm GeV} \simlt M_{\rm mess} \simlt 10^{13}~{\rm GeV}.
\label{eq:Mmess-range}
\end{equation}
Here, we have taken into account the existence of possible $O(1)$ 
factors to derive these numbers.

There are only two nontrivial phases appearing in the superparticle 
masses: the phase of $\mu$ in Eq.~(\ref{eq:mu-mH2}) and that of the 
gaugino masses in Eq.~(\ref{eq:gauginos}).  These can be absorbed into 
the phases of the fields using $R$ and Peccei-Quinn rotations, so that 
the supersymmetric $CP$ problem is absent.  The squark and slepton 
masses receive dominant contributions from gauge mediation, which are 
flavor universal and thus do not lead to the supersymmetric flavor 
problem.  On the other hand, they also receive renormalization group 
contributions from $m_{H_u}^2$ and $m_{H_d}^2$ through the Yukawa 
couplings in the energy interval between $M_*$ and the weak scale. 
This may lead to nontrivial flavor violating processes at a level 
consistent with but close to the current experimental bounds.  The 
spectrum of the superparticles is essentially that of gauge mediation, 
although the third generation superparticle masses, as well as the 
Higgs soft masses, can have significant deviations from it due to 
renormalization group contributions from $m_{H_u}^2$ and $m_{H_d}^2$. 
These deviations are parameterized by two free parameters of the 
model: $m_{H_u}^2$ and $m_{H_d}^2$ at $M_*$.  The predictions of 
Eqs.~(\ref{eq:N_mess},~\ref{eq:Mmess-range}), however, can be tested 
without taking into account these effects by using the first two 
generation superparticle masses.  Note that there is no correction 
from the hidden sector dynamics~\cite{Cohen:2006qc}, since the 
$X$ field is extremely weakly coupled below the messenger scale 
$M_{\rm mess}$.

Independently from the parameters of the model, the gravitino mass 
is given by
\begin{equation}
  m_{3/2} \simeq \frac{F_X}{\sqrt{3} M_{\rm Pl}} 
    \approx (1~{\rm -}~10)~{\rm GeV},
\label{eq:m32}
\end{equation}
implying that the gravitino is the lightest supersymmetric particle. 
The next-to-lightest supersymmetric particle, which is mostly 
the right-handed stau, then decays into the gravitino with a 
lifetime of order $\tau_{\tilde{\tau}} \simeq 48\pi m_{3/2}^2 
M_{\rm Pl}^2/m_{\tilde{\tau}}^5 \approx (10^2$~--~$10^6)~{\rm sec}$. 
This leads to interesting phenomenology at future collider experiments.

\section{Effective Field Theory Analysis}
\label{sec:eft}

The analysis performed in the previous sections does not strictly follow 
the method of effective field theory, in which heavy degrees of freedom 
are integrated out in discussing the dynamics of low energy excitations. 
Here we discuss the dynamics of the model using effective field theories.

For large values of $X$, $|X| \simgt \Lambda/\lambda$, the largest 
physical scale below the cutoff $M_*$ is the mass of the $F$, $\bar{F}$ 
fields, $|M_F| = |\lambda X|$.  Below this scale, the $F$, $\bar{F}$ 
fields are integrated out, leaving an effective field theory that 
contains only the $SU(5)_{\rm hid}$ gauge supermultiplet, MSSM fields, 
and $X$.  The integration generates several important operators. 
First, it generates
\begin{equation}
  {\cal L} = \left\{ -\int\!d^2\theta \sum_a 
    \frac{n_G}{32\pi^2} \left( \ln\frac{X}{|\langle X \rangle|} \right) 
    {\cal W}_a^\alpha {\cal W}_{a \alpha} + {\rm h.c.} \right\} 
  - \int\!d^4\theta \sum_\Phi \sum_a \frac{g_a^4 n_G}{(16\pi^2)^2} 
    C_a^\Phi \left( \ln\frac{X^\dagger X}{|\langle X \rangle|^2} \right)^2 
    \Phi^\dagger \Phi,
\label{eq:eft-gmsb}
\end{equation}
where ${\cal W}^a_\alpha$ $(a=1,2,3)$ are the field-strength superfields 
for $G_{\rm SM}$, and $\Phi = Q,U,D,L,E,H_u, H_d$ are the MSSM matter 
and Higgs superfields.  The couplings $g_a$ are evaluated at $|\lambda 
\langle X \rangle|$, and $\langle X \rangle$ will be determined by 
minimizing the potential in the low energy theory.  These operators 
become gauge-mediated gaugino and scalar masses when $\langle X \rangle 
\neq 0$ and $F_X \neq 0$~\cite{Giudice:1997ni}.  The operator
\begin{equation}
  {\cal L} = -\int\!d^2\theta 
    \frac{n_G}{32\pi^2} \left( \ln\frac{X}{|\langle X \rangle|} \right) 
    {\cal W}^\alpha {\cal W}_\alpha + {\rm h.c.},
\label{eq:eft-SU5}
\end{equation}
is also generated, where ${\cal W}_\alpha$ is the field-strength 
superfield for $SU(5)_{\rm hid}$.  Here, we have neglected the contribution 
from Eq.~(\ref{eq:W_H-X}), assuming that $\eta$ is sufficiently small 
($\eta \simlt \lambda$) so that it gives only a phenomenologically 
irrelevant effect.  Note that the first terms of Eq.~(\ref{eq:eft-gmsb}) 
and Eq.~(\ref{eq:eft-SU5}) are the manifestations of $U(1)_H$ mixed 
anomalies with $G_{\rm SM}$ and $SU(5)_{\rm hid}$, respectively.

Integrating out $F, \bar{F}$ also generates corrections to the K\"ahler 
potential containing $X$ and ${\cal W}_\alpha$.  Together with the 
tree-level terms in Eq.~(\ref{eq:K_X}), the K\"ahler potential for 
$X$ and ${\cal W}_\alpha$ is given by
\begin{equation}
  K = X^\dagger X - \frac{1}{4 M_*^2} (X^\dagger X)^2 
    + \delta K_{\rm CW} + \delta K_{{\cal W}^4} + \cdots,
\label{eq:eft-K}
\end{equation}
where $\delta K_{\rm CW}$ and $\delta K_{{\cal W}^4}$ are given above 
Eqs.~(\ref{eq:stab-cond}) and (\ref{eq:stab-cond-2}), respectively. 
(These terms are not important for the dynamics of the model in the 
parameter region we are interested.)  The effective theory below 
$|\lambda \langle X \rangle|$ is then given by Eqs.~(\ref{eq:eft-gmsb},%
~\ref{eq:eft-SU5},~\ref{eq:eft-K}), together with Eq.~(\ref{eq:K_H-X}), 
the MSSM kinetic and Yukawa terms, the gauge kinetic term for 
$SU(5)_{\rm hid}$, and the constant term in the superpotential. 
Since $X$ has only irrelevant interactions below $|\lambda \langle 
X \rangle|$, the operators of Eqs.~(\ref{eq:K_H-X},~\ref{eq:eft-gmsb}) 
run only by loops of the MSSM states.  As a result, renormalization 
group evolutions for $\mu$ and the supersymmetry breaking 
masses are exactly those of the MSSM below the messenger scale 
$M_{\rm mess} = |\lambda \langle X \rangle|$.

At the scale $|\Lambda_{\rm eff}| = |\lambda X \Lambda^2|^{1/3}$, 
$SU(5)_{\rm hid}$ gauge interactions become strong, giving nonperturbative 
effects.  The $SU(5)_{\rm hid}$ gauge multiplet should be integrated 
out.  In particular, the $SU(5)_{\rm hid}$ gauge kinetic term and 
Eq.~(\ref{eq:eft-SU5}) are replaced by the superpotential term of 
$\Lambda_{\rm eff}^3$, leading to the superpotential
\begin{equation}
  W = \lambda X \Lambda^2 + c,
\label{eq:eft-W}
\end{equation}
in the effective theory below $\Lambda_{\rm eff}$.  (The MSSM 
Yukawa terms should also exist.)  The combination ${\cal W}^\alpha 
{\cal W}_\alpha$ in Eq.~(\ref{eq:eft-K}) is also replaced by the 
condensation $\langle {\cal W}^\alpha {\cal W}_\alpha \rangle = 32\pi^2 
\lambda X \Lambda^2/n_G$, and the K\"ahler potential term
\begin{equation}
  \delta K_{\rm np} \approx |\Lambda_{\rm eff}|^2 
    \approx |\lambda \Lambda^2 X|^{2/3},
\label{eq:delta-K_np}
\end{equation}
is generated.  The theory now contains only the MSSM states and $X$, 
whose interactions are given by Eqs.~(\ref{eq:K_H-X},~\ref{eq:eft-gmsb},%
~\ref{eq:eft-K},~\ref{eq:eft-W},~\ref{eq:delta-K_np}) and the MSSM 
Yukawa couplings.

Below $|\Lambda_{\rm eff}| = |\lambda X \Lambda^2|^{1/3}$, the dynamics 
of $X$ decouple from the rest, so that the minimum of the potential, 
$\langle X \rangle$, is determined by Eqs.~(\ref{eq:eft-K},~\ref{eq:eft-W},%
~\ref{eq:delta-K_np}).  In order for the analysis to be consistent, the 
resulting $\langle X \rangle$ should satisfy $\Lambda/\lambda \simlt 
|\langle X \rangle| \simlt M_*$.  In fact, for $10^{-6} \simlt \lambda^2 
\simlt 10^{-3}$, the minimum is given by Eq.~(\ref{eq:X-FX-VEV}) and 
is within this range.  It is instructive to write $\langle X \rangle$ 
in the form
\begin{equation}
  \langle X \rangle 
    = \frac{2c(\lambda \Lambda^2)^\dagger}{|\lambda \Lambda^2|^2} 
      \frac{M_*^2}{M_{\rm Pl}^2},
\label{eq:X-VEV}
\end{equation}
although $\lambda \Lambda^2$ can be chosen real, and $c$ is set to 
$c = \lambda \Lambda^2 M_{\rm Pl}/\sqrt{3}$ by the condition of vanishing 
cosmological constant.  This shows why $X$ can obtain a nonzero VEV 
despite the fact that it is charged under both $U(1)_H$ and an accidental 
$U(1)_R$ symmetry possessed by the $\lambda$ coupling: $R(X) = 2$, 
$R(F) = R(\bar{F}) = 0$.  The $U(1)_H$ symmetry is broken by the anomaly, 
i.e., $\Lambda$ has a charge of $-1$, and $U(1)_R$ by the constant term 
in the superpotential, i.e., $c$ has a charge of $+2$.  The expression 
of Eq.~(\ref{eq:X-VEV}) respects both of these spurious symmetries. 
The mass of $X$ is given by Eq.~(\ref{eq:mX2}), which is much smaller 
than $\Lambda_{\rm eff}$.  Minimizing the potential in this low energy 
effective theory, therefore, is appropriate.

Finally, the expectation values of Eq.~(\ref{eq:X-FX-VEV}) give 
$\mu$ and the supersymmetry breaking masses through the operators 
Eqs.~(\ref{eq:K_H-X},~\ref{eq:eft-gmsb}).  The coefficients of the 
operators in Eq.~(\ref{eq:K_H-X}) (Eq.~(\ref{eq:eft-gmsb})) are 
subject to renormalization group evolution from $M_*$ ($|\lambda 
\langle X \rangle|$) to the scale of the superparticle masses caused 
by loops of the MSSM states.  The loop effects from $X,F,\bar{F}$ on 
Eq.~(\ref{eq:K_H-X}) between $M_*$ and $|\lambda \langle X \rangle|$ 
are negligible because of the small value of $\lambda$.

\section{More on {\boldmath $U(1)_H$}}
\label{sec:pheno}

The $U(1)_H$ charge assignment of Eqs.~(\ref{eq:H-XFF}~--~\ref{eq:H-Higgs}) 
contains two free parameters $x$ and $y$.  These parameters can be 
restricted by imposing various phenomenological requirements.  For 
example, if we require that dangerous dimension-five proton decay 
operators $W \sim QQQL$ and $UUDE$ are prohibited by $U(1)_H$, then 
we obtain the conditions $3x+y \neq 0$ and $3x+y \neq -4$, respectively. 
Similarly, if we require that $U(1)_H$ forbids dimension-four $R$-parity 
violating operators $W \sim LH_u$, $QDL$, $UDD$, $LLE$ and $K \sim 
L^\dagger H_d$, we obtain $y \neq -1$, $y \neq 1$, $x \neq -1$, 
$y \neq 1$ and $y \neq 1$.  If the values of $x$ and $y$ satisfy 
these conditions, therefore, sufficient proton stability is ensured.%
\footnote{Proton decay at a dangerous level may still be caused 
 by operators $W \sim X^m QQQL$ and/or $X^m UUDE$ ($m \geq 1$) through 
 the $X$ VEV.  The absence of these operators, however, is consistent 
 with the absence of the operators $K \sim X^\dagger X M^\dagger M$ 
 ($M = Q,U,D,L,E$), which we have assumed.  Alternatively, the 
 coefficients of these operators, even if they exist, may be suppressed 
 by the corresponding Yukawa coupling factors, in which case the 
 proton is sufficiently stable.  Yet another possibility is to 
 impose $3x+y \neq -8,-6,-4,-2,0,2,4$, suppressing all the operators 
 $W \sim X^m QQQL$, $X^m UUDE$ with $m \leq 4$.}
The charge assignment of $x = y = -1/2$ was discussed in 
Ref.~\cite{Ibe:2007km}.

To generate light neutrino masses, we can introduce three generations 
of right-handed neutrino superfields $N$ with the Yukawa couplings 
$W \sim L N H_u$.  This determines their $U(1)_H$ charges to be
\begin{equation}
  N(-1-y).
\label{eq:H-N}
\end{equation}
An interesting possibility arises if $y=0$.  In this case the 
superpotential can have interactions of the form $XN^2$, so that 
the right-handed neutrinos can have the superpotential
\begin{equation}
  W = \frac{\kappa}{2} X N^2 + y_\nu L N H_u,
\label{eq:seesaw}
\end{equation}
where $\kappa$ and $y_\nu$ are $3 \times 3$ matrices in generation 
space.  With the $X$ VEV of Eq.~(\ref{eq:X-FX-VEV}), this generates 
small neutrino masses through the seesaw mechanism.  Note that the 
vacuum of Eq.~(\ref{eq:X-FX-VEV}) is not destabilized if $\kappa 
\simlt O(0.1)$, as can be seen from Eq.~(\ref{eq:stab-cond}) with 
$\lambda \rightarrow \kappa$ and $n_G^2 \rightarrow 3$.

A $U(1)_H$ charge assignment that satisfies all the requirements 
above can be obtained with
\begin{equation}
  x = \frac{4}{3} + 2n,
\qquad
  y = 0,
\label{eq:x-y}
\end{equation}
where $n$ is an integer.  The $U(1)_H$ symmetry is spontaneously broken 
by the VEV of $X$.  The charge assignment of Eq.~(\ref{eq:x-y}), however, 
leaves a discrete $Z_6$ symmetry after the breaking.  The product of 
$Z_6$ and $U(1)_Y$ contains the (anomalous) $Z_3$ baryon number and 
(anomaly-free) $Z_2$ matter parity ($R$ parity) as subgroups.  This 
symmetry thus strictly forbids the $R$-parity violating operators, 
and the lightest supersymmetric particle is absolutely stable.

We note that the requirements on $U(1)_H$ discussed above are not 
a necessity.  For example, the $R$-parity violating operators may be 
forbidden by imposing a matter (or $R$) parity in addition to $U(1)_H$, 
and the dimension-five proton decay operators may be suppressed by 
some mechanism in the ultraviolet theory.  Small neutrino masses 
may also be obtained by the seesaw mechanism with $y=-1$ or if there 
is additional $U(1)_H$ breaking, or they may simply arise from small 
Yukawa couplings without the Majorana mass terms for $N$.  Nevertheless, 
it is interesting that $U(1)_H$ can be used to address these issues.%
\footnote{An interesting possibility is to assign $U(1)_H$ charges 
 that depend on the generations, which allows us to use $U(1)_H$ 
 also as a $U(1)$ flavor symmetry.  For example, we can assign 
 the charges so that the light generation Yukawa couplings arise 
 only through powers of the $X$ VEV, e.g. $W \sim \langle X \rangle^m 
 QUH_u$~\cite{Froggatt:1978nt}.  It is, however, not clear if the 
 existence of such operators is consistent with the assumption that 
 the operators $K \sim X^\dagger X M^\dagger M$ ($M = Q,U,D,L,E$) 
 are absent.}

We finally discuss possible origins of $U(1)_H$.  Since any global 
symmetry is expected to be broken by quantum gravity effects, we may 
expect that the ultimate origin of $U(1)_H$ is a gauge symmetry.  One 
possibility is that $U(1)_H$ is a remnant of a pseudo-anomalous $U(1)$ 
gauge symmetry arising in string theory.  Another possibility is that 
the $U(1)_H$ symmetry is a gauge symmetry in higher dimensional spacetime 
broken on a ``distant brane.''  The superfields of Eqs.~(\ref{eq:H-XFF},%
~\ref{eq:H-matter},~\ref{eq:H-Higgs},~\ref{eq:H-N}) are localized 
on some brane, while the $G_{\rm SM}$ and $G_{\rm hid}$ gauge fields 
propagate in the extra dimensional bulk (which we assume to be small). 
The $U(1)_H$-$G_{\rm SM}$ and $U(1)_H$-$G_{\rm hid}$ anomalies of the 
original $U(1)_H$ gauge symmetry are canceled by particles $\psi$ on 
the ``distant brane,'' which become massive through $U(1)_H$ breaking 
there.  This effectively leaves an anomalous global $U(1)_H$ symmetry 
on ``our brane.''  The scale of $U(1)_H$ breaking $v_{\rm dist}$ on the 
distant brane should be higher than $M_F = \lambda \langle X \rangle$ 
so that the theory is reduced to the model presented here at low 
energies.  If $v_{\rm dist} \simlt M_*$, we have extra vector-like 
states $\psi$ with masses of order $v_{\rm dist}$, which are charged 
under $G_{\rm SM} \times G_{\rm hid}$ but have no superpotential 
interactions with the other fields of the model.  The existence of such 
states, however, does not destroy successful gauge coupling unification 
(as long as they are sufficiently heavy), since they are in complete 
representations of $SU(5)_{\rm SM} \supset G_{\rm SM}$.

\section{Summary and Discussions}
\label{sec:summary}

We have presented a simple and realistic model of supersymmetry breaking. 
The gauge group of the model is $SU(5)_{\rm hid} \times G_{\rm SM}$, 
and the matter content is given in Table~\ref{table:model}.
\begin{table}
\begin{center}
\begin{tabular}{c|cc|c}
  & $SU(5)_{\rm hid}$ & $SU(5)_{\rm SM} \supset G_{\rm SM}$ 
  & $U(1)_H$ \\ \hline
     $X$    &      ${\bf 1}$     &      ${\bf 1}$     &    $2$   \\
     $F$    &       $\fund$      & $\overline{\fund}$ &   $-1$   \\
  $\bar{F}$ & $\overline{\fund}$ &       $\fund$      &   $-1$   \\ \hline
    $H_u$   & ${\bf 1}$ &  $({\bf 1},{\bf 2})_{1/2}$  &    $1$   \\
    $H_d$   & ${\bf 1}$ &  $({\bf 1},{\bf 2})_{-1/2}$ &    $1$   \\
     $M$    & ${\bf 1}$ & $3 \times \Bigl( \overline{\fund}+\asym \Bigr)$ 
            & $-\frac{1}{2} + 3\Bigl(x+\frac{1}{2}\Bigr)q_B
               + \Bigl(y+\frac{1}{2}\Bigr)q_L$
\end{tabular}
\end{center}
\caption{The entire matter content of the model.  Here, $H_u$, $H_d$ 
 and $M$ represent the MSSM Higgs and matter fields, and $q_B$ and 
 $q_L$ the baryon and lepton numbers, respectively.}
\label{table:model}
\end{table}
The form of the superpotential and K\"ahler potential interactions 
are controlled by the (anomalous) global $U(1)_H$ symmetry, which can 
also be used to prohibit dangerous proton decay operators.  With the 
higher dimension operators suppressed by the cutoff scale $M_* \simeq 
10^{16}~{\rm GeV}$, the supersymmetry breaking masses and the $\mu$ 
parameter are generated with the same order of magnitude.  No flavor 
violating or $CP$ violating terms arise.

The model requires the absence of direct interactions between the 
supersymmetry breaking and matter fields suppressed by the cutoff 
scale $M_*$.  This should be understood as a property of the theory 
at or above $M_*$.  The model also requires the coefficient $\lambda$ 
of the superpotential interaction $XF\bar{F}$ to satisfy $\lambda^2 
\simlt 10^{-3}$.  (The coefficient $\eta$ of the interaction 
$F\bar{F}H_u H_d/M_*$ must also satisfy $\eta \simlt \lambda$.) 
In particular, in order for our present analysis of the dynamics 
to be valid, the coefficient must be in the range $10^{-6} \simlt 
\lambda^2 \simlt 10^{-3}$.  This range, however, is not that small. 
Moreover, since the bound arises from requiring the existence of 
a supersymmetry breaking minimum, a required value of $\lambda$ 
may arise naturally as a result of anthropic selection.

The model provides several definite predictions on the spectrum 
of superparticles.  The spectrum of the first two generation 
superparticles is that of minimal gauge mediation with the number 
of messengers $N_{\rm mess} = 5$ and the messenger scale $M_{\rm mess} 
\simlt 10^{13}~{\rm GeV}$.  The condition $M_{\rm mess} \simgt 
10^{11}~{\rm GeV}$ also arises if one focuses on the regime 
where the hidden sector gauge dynamics is perturbative at 
the messenger scale.%
\footnote{A consistent supersymmetry breaking minimum exists for 
 $10^{5}~{\rm GeV} \simlt M_{\rm mess} \simlt 10^{10}~{\rm GeV}$ 
 in the case where the hidden sector gauge dynamics becomes strongly 
 coupled above the messenger scale, as shown in Ref.~\cite{Ibe:2007gf}.}
On the other hand, the spectrum of the Higgs bosons and third 
generation superparticles can have deviations from that of minimal 
gauge mediation because of the tree-level contributions to the Higgs 
mass-squared parameters, $m_{H_u}^2$ and $m_{H_d}^2$, at $M_*$. 
The lightest supersymmetric particle is the gravitino with 
a mass of order $(1~{\rm -}~10)~{\rm GeV}$.

How robust are these predictions under modifications to the model? 
There are several levels of modifications one can consider.  For 
example, one can consider changing the hidden sector gauge group 
to $SU(N_{\rm hid})$ ($N_{\rm hid} > 5$), with $F$ and $\bar{F}$ 
transforming as ${\bf 5}^* + (N_{\rm hid}-1)\,{\bf 1}$ and 
${\bf 5} + (N_{\rm hid}-1)\,{\bf 1}$ under $SU(5)_{\rm SM}$. 
In this case, the prediction of $N_{\rm mess} = 5$ will be 
lost (it becomes $N_{\rm mess} > 5$), although the one for 
$M_{\rm mess}$ essentially remains (unless $N_{\rm hid}$ is 
much larger than $5$).  More drastic relaxations of the predictions, 
however, could also occur if we replace $F$ and $\bar{F}$ by 
$F_{\rm hid}({\bf N_{\rm hid}}, {\bf 1}) + F_{\rm SM}({\bf 1}, 
{\bf r}^*)$ and $\bar{F}_{\rm hid}({\bf N_{\rm hid}}^*, {\bf 1}) 
+ \bar{F}_{\rm SM}({\bf 1}, {\bf r})$, respectively, where the 
numbers in parentheses represent the transformation properties 
under $SU(N_{\rm hid}) \times SU(5)_{\rm SM}$.  Here, $N_{\rm hid}$ 
is an arbitrary integer larger than $1$, and ${\bf r}$ is an 
arbitrary (in general reducible) representation of $SU(5)_{\rm SM}$. 
In this case, the couplings for the $X F_{\rm hid} \bar{F}_{\rm hid}$ 
and $X F_{\rm SM} \bar{F}_{\rm SM}$ interactions can differ, 
and the only remaining prediction on $N_{\rm mess}$ and 
$M_{\rm mess}$ is $M_{\rm mess} \simlt 10^{13}~{\rm GeV}$, 
which is obtained by Eq.~(\ref{eq:stab-cond}) with $n_G^2 
\rightarrow {\rm dim}({\bf r})$.  The gravitino mass, 
however, is still of order $(1~{\rm -}~10)~{\rm GeV}$.

As the LHC turns on, we will start acquiring the data on the masses 
and decay modes of the supersymmetric particles if weak scale 
supersymmetry is realized in nature.  It will then be interesting 
to see if this data is consistent with (one of) the pattern(s) 
discussed in this paper.  Such an exploration may shed some light 
on the origin of supersymmetry breaking, including the structure, 
e.g., the gauge group and matter content, of the hidden sector.

\section*{Acknowledgment}

This work was supported in part by the U.S. DOE under Contract 
DE-AC03-76SF00098, and in part by the NSF under grant PHY-04-57315. 
The work of Y.N. was also supported by the NSF under grant PHY-0555661, 
by a DOE OJI, and by an Alfred P. Sloan Research Foundation. 
M.P. thanks the Aspen Center for Physics, where a part of this 
work was performed.

\end{document}

%% file: figure.tex
\setlength{\unitlength}{0.240900pt}
\ifx\plotpoint\undefined\newsavebox{\plotpoint}\fi
\sbox{\plotpoint}{\rule[-0.200pt]{0.400pt}{0.400pt}}%
\begin{picture}(1500,900)(0,0)
\font\gnuplot=cmtt10 at 10pt
\gnuplot
\sbox{\plotpoint}{\rule[-0.200pt]{0.400pt}{0.400pt}}%
\sbox{\plotpoint}{\rule[-0.600pt]{1.200pt}{1.200pt}}%
\put(180,606){\usebox{\plotpoint}}
\put(180.0,606.0){\rule[-0.600pt]{274.626pt}{1.200pt}}
\sbox{\plotpoint}{\rule[-0.200pt]{0.400pt}{0.400pt}}%
\put(180,151){\usebox{\plotpoint}}
\multiput(180.58,151.00)(0.492,0.712){21}{\rule{0.119pt}{0.667pt}}
\multiput(179.17,151.00)(12.000,15.616){2}{\rule{0.400pt}{0.333pt}}
\multiput(192.58,168.00)(0.492,2.430){19}{\rule{0.118pt}{1.991pt}}
\multiput(191.17,168.00)(11.000,47.868){2}{\rule{0.400pt}{0.995pt}}
\multiput(203.58,220.00)(0.492,3.340){21}{\rule{0.119pt}{2.700pt}}
\multiput(202.17,220.00)(12.000,72.396){2}{\rule{0.400pt}{1.350pt}}
\multiput(215.58,298.00)(0.492,4.222){19}{\rule{0.118pt}{3.373pt}}
\multiput(214.17,298.00)(11.000,83.000){2}{\rule{0.400pt}{1.686pt}}
\multiput(226.58,388.00)(0.492,3.770){21}{\rule{0.119pt}{3.033pt}}
\multiput(225.17,388.00)(12.000,81.704){2}{\rule{0.400pt}{1.517pt}}
\multiput(238.58,476.00)(0.492,3.467){19}{\rule{0.118pt}{2.791pt}}
\multiput(237.17,476.00)(11.000,68.207){2}{\rule{0.400pt}{1.395pt}}
\multiput(249.58,550.00)(0.492,2.349){21}{\rule{0.119pt}{1.933pt}}
\multiput(248.17,550.00)(12.000,50.987){2}{\rule{0.400pt}{0.967pt}}
\multiput(261.58,605.00)(0.492,1.628){19}{\rule{0.118pt}{1.373pt}}
\multiput(260.17,605.00)(11.000,32.151){2}{\rule{0.400pt}{0.686pt}}
\multiput(272.58,640.00)(0.492,0.755){21}{\rule{0.119pt}{0.700pt}}
\multiput(271.17,640.00)(12.000,16.547){2}{\rule{0.400pt}{0.350pt}}
\multiput(284.00,658.59)(0.943,0.482){9}{\rule{0.833pt}{0.116pt}}
\multiput(284.00,657.17)(9.270,6.000){2}{\rule{0.417pt}{0.400pt}}
\multiput(307.00,662.95)(2.248,-0.447){3}{\rule{1.567pt}{0.108pt}}
\multiput(307.00,663.17)(7.748,-3.000){2}{\rule{0.783pt}{0.400pt}}
\multiput(318.00,659.95)(2.472,-0.447){3}{\rule{1.700pt}{0.108pt}}
\multiput(318.00,660.17)(8.472,-3.000){2}{\rule{0.850pt}{0.400pt}}
\multiput(330.00,656.95)(2.248,-0.447){3}{\rule{1.567pt}{0.108pt}}
\multiput(330.00,657.17)(7.748,-3.000){2}{\rule{0.783pt}{0.400pt}}
\multiput(341.00,653.95)(2.472,-0.447){3}{\rule{1.700pt}{0.108pt}}
\multiput(341.00,654.17)(8.472,-3.000){2}{\rule{0.850pt}{0.400pt}}
\multiput(353.00,650.95)(2.248,-0.447){3}{\rule{1.567pt}{0.108pt}}
\multiput(353.00,651.17)(7.748,-3.000){2}{\rule{0.783pt}{0.400pt}}
\put(364,647.17){\rule{2.500pt}{0.400pt}}
\multiput(364.00,648.17)(6.811,-2.000){2}{\rule{1.250pt}{0.400pt}}
\put(376,645.17){\rule{2.300pt}{0.400pt}}
\multiput(376.00,646.17)(6.226,-2.000){2}{\rule{1.150pt}{0.400pt}}
\put(387,643.17){\rule{2.500pt}{0.400pt}}
\multiput(387.00,644.17)(6.811,-2.000){2}{\rule{1.250pt}{0.400pt}}
\put(399,641.17){\rule{2.300pt}{0.400pt}}
\multiput(399.00,642.17)(6.226,-2.000){2}{\rule{1.150pt}{0.400pt}}
\put(410,639.17){\rule{2.500pt}{0.400pt}}
\multiput(410.00,640.17)(6.811,-2.000){2}{\rule{1.250pt}{0.400pt}}
\put(422,637.17){\rule{2.300pt}{0.400pt}}
\multiput(422.00,638.17)(6.226,-2.000){2}{\rule{1.150pt}{0.400pt}}
\put(433,635.67){\rule{2.891pt}{0.400pt}}
\multiput(433.00,636.17)(6.000,-1.000){2}{\rule{1.445pt}{0.400pt}}
\put(445,634.17){\rule{2.300pt}{0.400pt}}
\multiput(445.00,635.17)(6.226,-2.000){2}{\rule{1.150pt}{0.400pt}}
\put(456,632.17){\rule{2.500pt}{0.400pt}}
\multiput(456.00,633.17)(6.811,-2.000){2}{\rule{1.250pt}{0.400pt}}
\put(468,630.67){\rule{2.650pt}{0.400pt}}
\multiput(468.00,631.17)(5.500,-1.000){2}{\rule{1.325pt}{0.400pt}}
\put(479,629.17){\rule{2.500pt}{0.400pt}}
\multiput(479.00,630.17)(6.811,-2.000){2}{\rule{1.250pt}{0.400pt}}
\put(491,627.67){\rule{2.650pt}{0.400pt}}
\multiput(491.00,628.17)(5.500,-1.000){2}{\rule{1.325pt}{0.400pt}}
\put(502,626.17){\rule{2.500pt}{0.400pt}}
\multiput(502.00,627.17)(6.811,-2.000){2}{\rule{1.250pt}{0.400pt}}
\put(514,624.67){\rule{2.650pt}{0.400pt}}
\multiput(514.00,625.17)(5.500,-1.000){2}{\rule{1.325pt}{0.400pt}}
\put(525,623.17){\rule{2.500pt}{0.400pt}}
\multiput(525.00,624.17)(6.811,-2.000){2}{\rule{1.250pt}{0.400pt}}
\put(537,621.67){\rule{2.650pt}{0.400pt}}
\multiput(537.00,622.17)(5.500,-1.000){2}{\rule{1.325pt}{0.400pt}}
\put(548,620.67){\rule{2.891pt}{0.400pt}}
\multiput(548.00,621.17)(6.000,-1.000){2}{\rule{1.445pt}{0.400pt}}
\put(560,619.67){\rule{2.891pt}{0.400pt}}
\multiput(560.00,620.17)(6.000,-1.000){2}{\rule{1.445pt}{0.400pt}}
\put(572,618.17){\rule{2.300pt}{0.400pt}}
\multiput(572.00,619.17)(6.226,-2.000){2}{\rule{1.150pt}{0.400pt}}
\put(583,616.67){\rule{2.891pt}{0.400pt}}
\multiput(583.00,617.17)(6.000,-1.000){2}{\rule{1.445pt}{0.400pt}}
\put(595,615.67){\rule{2.650pt}{0.400pt}}
\multiput(595.00,616.17)(5.500,-1.000){2}{\rule{1.325pt}{0.400pt}}
\put(606,614.67){\rule{2.891pt}{0.400pt}}
\multiput(606.00,615.17)(6.000,-1.000){2}{\rule{1.445pt}{0.400pt}}
\put(618,613.67){\rule{2.650pt}{0.400pt}}
\multiput(618.00,614.17)(5.500,-1.000){2}{\rule{1.325pt}{0.400pt}}
\put(629,612.67){\rule{2.891pt}{0.400pt}}
\multiput(629.00,613.17)(6.000,-1.000){2}{\rule{1.445pt}{0.400pt}}
\put(295.0,664.0){\rule[-0.200pt]{2.891pt}{0.400pt}}
\put(652,611.67){\rule{2.891pt}{0.400pt}}
\multiput(652.00,612.17)(6.000,-1.000){2}{\rule{1.445pt}{0.400pt}}
\put(664,610.67){\rule{2.650pt}{0.400pt}}
\multiput(664.00,611.17)(5.500,-1.000){2}{\rule{1.325pt}{0.400pt}}
\put(675,609.67){\rule{2.891pt}{0.400pt}}
\multiput(675.00,610.17)(6.000,-1.000){2}{\rule{1.445pt}{0.400pt}}
\put(641.0,613.0){\rule[-0.200pt]{2.650pt}{0.400pt}}
\put(698,608.67){\rule{2.891pt}{0.400pt}}
\multiput(698.00,609.17)(6.000,-1.000){2}{\rule{1.445pt}{0.400pt}}
\put(710,607.67){\rule{2.650pt}{0.400pt}}
\multiput(710.00,608.17)(5.500,-1.000){2}{\rule{1.325pt}{0.400pt}}
\put(687.0,610.0){\rule[-0.200pt]{2.650pt}{0.400pt}}
\put(733,606.67){\rule{2.650pt}{0.400pt}}
\multiput(733.00,607.17)(5.500,-1.000){2}{\rule{1.325pt}{0.400pt}}
\put(721.0,608.0){\rule[-0.200pt]{2.891pt}{0.400pt}}
\put(767,605.67){\rule{2.891pt}{0.400pt}}
\multiput(767.00,606.17)(6.000,-1.000){2}{\rule{1.445pt}{0.400pt}}
\put(744.0,607.0){\rule[-0.200pt]{5.541pt}{0.400pt}}
\put(917,605.67){\rule{2.650pt}{0.400pt}}
\multiput(917.00,605.17)(5.500,1.000){2}{\rule{1.325pt}{0.400pt}}
\put(779.0,606.0){\rule[-0.200pt]{33.244pt}{0.400pt}}
\put(940,606.67){\rule{2.891pt}{0.400pt}}
\multiput(940.00,606.17)(6.000,1.000){2}{\rule{1.445pt}{0.400pt}}
\put(928.0,607.0){\rule[-0.200pt]{2.891pt}{0.400pt}}
\put(963,607.67){\rule{2.891pt}{0.400pt}}
\multiput(963.00,607.17)(6.000,1.000){2}{\rule{1.445pt}{0.400pt}}
\put(952.0,608.0){\rule[-0.200pt]{2.650pt}{0.400pt}}
\put(986,608.67){\rule{2.891pt}{0.400pt}}
\multiput(986.00,608.17)(6.000,1.000){2}{\rule{1.445pt}{0.400pt}}
\put(998,609.67){\rule{2.650pt}{0.400pt}}
\multiput(998.00,609.17)(5.500,1.000){2}{\rule{1.325pt}{0.400pt}}
\put(975.0,609.0){\rule[-0.200pt]{2.650pt}{0.400pt}}
\put(1021,610.67){\rule{2.650pt}{0.400pt}}
\multiput(1021.00,610.17)(5.500,1.000){2}{\rule{1.325pt}{0.400pt}}
\put(1032,611.67){\rule{2.891pt}{0.400pt}}
\multiput(1032.00,611.17)(6.000,1.000){2}{\rule{1.445pt}{0.400pt}}
\put(1044,612.67){\rule{2.650pt}{0.400pt}}
\multiput(1044.00,612.17)(5.500,1.000){2}{\rule{1.325pt}{0.400pt}}
\put(1055,613.67){\rule{2.891pt}{0.400pt}}
\multiput(1055.00,613.17)(6.000,1.000){2}{\rule{1.445pt}{0.400pt}}
\put(1067,614.67){\rule{2.650pt}{0.400pt}}
\multiput(1067.00,614.17)(5.500,1.000){2}{\rule{1.325pt}{0.400pt}}
\put(1078,615.67){\rule{2.891pt}{0.400pt}}
\multiput(1078.00,615.17)(6.000,1.000){2}{\rule{1.445pt}{0.400pt}}
\put(1090,616.67){\rule{2.650pt}{0.400pt}}
\multiput(1090.00,616.17)(5.500,1.000){2}{\rule{1.325pt}{0.400pt}}
\put(1101,617.67){\rule{2.891pt}{0.400pt}}
\multiput(1101.00,617.17)(6.000,1.000){2}{\rule{1.445pt}{0.400pt}}
\put(1113,618.67){\rule{2.650pt}{0.400pt}}
\multiput(1113.00,618.17)(5.500,1.000){2}{\rule{1.325pt}{0.400pt}}
\put(1124,619.67){\rule{2.891pt}{0.400pt}}
\multiput(1124.00,619.17)(6.000,1.000){2}{\rule{1.445pt}{0.400pt}}
\put(1136,621.17){\rule{2.300pt}{0.400pt}}
\multiput(1136.00,620.17)(6.226,2.000){2}{\rule{1.150pt}{0.400pt}}
\put(1147,622.67){\rule{2.891pt}{0.400pt}}
\multiput(1147.00,622.17)(6.000,1.000){2}{\rule{1.445pt}{0.400pt}}
\put(1159,623.67){\rule{2.650pt}{0.400pt}}
\multiput(1159.00,623.17)(5.500,1.000){2}{\rule{1.325pt}{0.400pt}}
\put(1170,625.17){\rule{2.500pt}{0.400pt}}
\multiput(1170.00,624.17)(6.811,2.000){2}{\rule{1.250pt}{0.400pt}}
\put(1182,626.67){\rule{2.650pt}{0.400pt}}
\multiput(1182.00,626.17)(5.500,1.000){2}{\rule{1.325pt}{0.400pt}}
\put(1193,628.17){\rule{2.500pt}{0.400pt}}
\multiput(1193.00,627.17)(6.811,2.000){2}{\rule{1.250pt}{0.400pt}}
\put(1205,629.67){\rule{2.650pt}{0.400pt}}
\multiput(1205.00,629.17)(5.500,1.000){2}{\rule{1.325pt}{0.400pt}}
\put(1216,631.17){\rule{2.500pt}{0.400pt}}
\multiput(1216.00,630.17)(6.811,2.000){2}{\rule{1.250pt}{0.400pt}}
\put(1228,633.17){\rule{2.300pt}{0.400pt}}
\multiput(1228.00,632.17)(6.226,2.000){2}{\rule{1.150pt}{0.400pt}}
\put(1239,634.67){\rule{2.891pt}{0.400pt}}
\multiput(1239.00,634.17)(6.000,1.000){2}{\rule{1.445pt}{0.400pt}}
\put(1251,636.17){\rule{2.300pt}{0.400pt}}
\multiput(1251.00,635.17)(6.226,2.000){2}{\rule{1.150pt}{0.400pt}}
\put(1262,638.17){\rule{2.500pt}{0.400pt}}
\multiput(1262.00,637.17)(6.811,2.000){2}{\rule{1.250pt}{0.400pt}}
\put(1274,640.17){\rule{2.300pt}{0.400pt}}
\multiput(1274.00,639.17)(6.226,2.000){2}{\rule{1.150pt}{0.400pt}}
\put(1285,642.17){\rule{2.500pt}{0.400pt}}
\multiput(1285.00,641.17)(6.811,2.000){2}{\rule{1.250pt}{0.400pt}}
\put(1297,644.17){\rule{2.300pt}{0.400pt}}
\multiput(1297.00,643.17)(6.226,2.000){2}{\rule{1.150pt}{0.400pt}}
\put(1308,646.17){\rule{2.500pt}{0.400pt}}
\multiput(1308.00,645.17)(6.811,2.000){2}{\rule{1.250pt}{0.400pt}}
\put(1009.0,611.0){\rule[-0.200pt]{2.891pt}{0.400pt}}
\put(90,846){\makebox(0,0){\large $V(X)$}}
\put(1315,560){\makebox(0,0){\large $X$}}
\put(845,618){\makebox(0,0){\Large $\bullet$}}
\put(845,555){\makebox(0,0){\large $\langle X \rangle$}}
\sbox{\plotpoint}{\rule[-0.600pt]{1.200pt}{1.200pt}}%
\put(180,40){\usebox{\plotpoint}}
\put(180.0,40.0){\rule[-0.600pt]{1.200pt}{197.538pt}}
\end{picture}